\documentclass[12pt]{article}

\usepackage{scicite}

\usepackage{times}

\usepackage{xcolor}
\usepackage{soul}

\newenvironment{sciabstract}{%
\begin{quote} \bf}
{\end{quote}}

\usepackage{graphicx}

\title{Multi-octave, CEP-stable source for high-energy field synthesis}

\author
{Ayman Alismail,$^{1,2+}$ Haochuan Wang,$^{1,3+}$ Gaia Barbiero,$^{1,3}$\\ Najd Altwaijry,$^{1,3}$ Syed Ali Hussain,$^{1,3}$ Volodymyr Pervak,$^{1}$\\ Wolfgang Schweinberger,$^{1}$ Abdallah M. Azzeer,$^{2}$\\ Ferenc Krausz,$^{1,3}$ Hanieh Fattahi,$^{1,3\ast}$\\
\\
\normalsize{$^{1}$Ludwig-Maximilians-University of Munich, Faculty of Physics,}\\
\normalsize{Am Coulombwall 1, 85748 Garching, Germany}\\
\normalsize{$^{2}$Physics and Astronomy Department, College of Sciences, King Saud University,}\\
\normalsize{Riyadh 11451, Saudi Arabia}\\
\normalsize{$^{3}$Max Planck Institute of Quantum Optics,}\\
\normalsize{Hans-Kopfermann-Str. 1, 85748 Garching, Germany}\\
\\
\normalsize{$^\ast$To whom correspondence should be addressed; E-mail:  hanieh.fattahi@mpq.mpg.de.}
\\
\normalsize{$^+$These authors contributed equally to this work.}
}

\date{}

\begin{document} 


\baselineskip24pt


\maketitle 


\begin{sciabstract}
The development of high-energy, high-power, multi-octave light-transients is currently subject of intense research driven by emerging applications in attosecond spectroscopy and coherent control. We report on a phase-stable, multi-octave source based on a Yb:YAG amplifier for light-transient generation. We demonstrate the amplification of a two-octave spectrum to 25\,$\mu$J of energy in two broadband amplification channels and their temporal compression to 6\,fs and 18\,fs at 1\,$\mu$m and 2\,$\mu$m, respectively. In this scheme due to the intrinsic temporal synchronization between the pump and seed pulses, the temporal jitter is restricted to long-term drift. We show that the intrinsic stability of the synthesizer allows for sub-cycle detection of an electric field at 0.15\,PHz. The complex electric field of the 0.15\,PHz pulses and their free-induction decay after interaction with water molecules are resolved by electro-optic sampling over 2\,ps. The scheme is scalable in peak- and average-power.
\end{sciabstract}


\subsection*{Teaser}
We report on a novel source for generating high energy, sub-cycle pulses based on a Yb thin-disk laser.

\section*{Introduction}

Controlling, shaping, and measuring the electric field of light at terahertz (THz) frequencies opened new possibilities for variety of applications from fundamental science to real world applications \cite{Dhillon2017}. Their extension to petahertz (PHz) frequencies has enhanced the depth of our insight into microscopic ultrafast dynamics at femtosecond and attosecond time scale. It was the controlling of the slippage of the electric field of light relative to its envelope (CEP-stability) that paved the way for isolated attosecond pulse generation \cite{Hentschel2001}. Nowadays, attosecond XUV pulses are generated routinely in laboratories and allow for control and observation of electron dynamics \cite{Krausz2014,Kim2013,Sommer2016,Schultze2010d,Drescher2002}. Nevertheless, attosecond metrology and spectroscopy has been restricted to the use of low-energy attosecond XUV pulses in combination with strong few-cycle near-infrared fields \cite{Leone2014}. Pushing the frontiers of attosecond spectroscopy to new areas like XUV-pump/XUV-probe spectroscopy \cite{Takahashi2013, Tzallas2011} or attosecond X-ray diffraction \cite{Fattahi2016f} to capture the four-dimensional microscopic dynamics of electrons outside the atomic core \cite{Shorokhov2016} calls for isolated XUV or X-ray attosecond pulses, with sufficient yield. Recently, there has been growing interest in the development of high energy, long wavelength, few-cycle lasers \cite{Sanchez2016a,Malevich2016}, as the cutoff energy in high-order harmonic generation (HHG) scales linearly with the peak intensity and quadratically with the wavelength of the driving pulse. However, the harmonic yield at longer wavelength is dramatically reduced, due to the higher quantum diffusion and lower recombination probability \cite{Popmintchev2015,Gordon2005,Popmintchev2012b,Teichmann2016}.

In parallel, many research studies have focused on enhancing the photon flux in HHG by optimizing the shape of the driving field \cite{Chipperfield2009,Jin2015a,Haessler2014}. It has been theoretically shown that light transients are superior to few-cycle pulses in pushing the frontiers of HHG towards X-rays with higher harmonics yield \cite{Kim2013,Moulet2014,Wendl2018}. Here, the presence of the longer wavelengths in the driving field furnishes the recolliding electrons with more energy. As the multi-octave field is confined to a single field cycle, the pre-ionization of the atoms is reduced \cite{Hassan2016}. Moreover, tailoring the spectral chirp of the light transient allows for better control of tunneling ionization, and subsequent electronic motion.

Light transients with tens of microjoules of energy have been generated by spectral shaping, or field synthesis of multi-octave pulses \cite{Cundiff2010,Wirth2013,Krauss2010}. However, increasing their energy to multi-mJ at several kHz repetition rates calls for a power- and energy-scalable scheme \cite{Mucke2015,Fattahi2014}. Optical parametric chirped-pulse amplifiers (OPCPA) are capable of delivering few-cycle pulses, scalable in peak- and average-power. Therefore, coherent combination of few-cycle pulses from multiple broadband OPCPAs allows for the generation of light transients with a higher peak- and average-power \cite{Manzoni2012,Huang2011,Mucke2015,Fattahi2014}. Here, the energy of few-cycle pulses at different central frequencies is boosted in OPCPA chains prior to the synthesis of super-octave waveforms via phase-coherent superposition of few-cycle pulses of different carrier frequency. Realizing such a concept requires high-energy lasers to pump and seed several OPCPA chains at different carrier frequencies. Direct generation of the multi-octave, CEP-stable seed spectrum from the pump laser allows for intrinsic temporal synchronization between: i) the pump and the seed pulses at the OPCPA stages, and ii) different OPCPA channels.

Based on this scheme, in 2018, Rossi et al., boosted the energy of the light transients to nearly 1\,mJ by employing a 20\,mJ, 1\,kHz, 150\,fs Ti:Sa amplifier to: i) generate multi-octave seed pulses and ii) pump the two parallel OPCPA channels \cite{MariaRossi2018}. In a follow-up experiment, Mainz et al., demonstrated the effect of the shape of the generated light transients on HHG yield \cite{Mainz2018}. However, boosting the energy and average power of light transients beyond this value calls for a different frontend due to the limitations of simultaneous energy and power scalability in Ti:Sa amplifiers \cite{Fattahi2015a}.

Yb:YAG lasers in thin-disk and slab geometries are capable of delivering pulses at high energy and average power simultaneously \cite{Russbueldt2010a,Zapata2016,nubbemeyer2017a}. Among two, the simultaneous scaling is more optimum in thin-disk lasers, due to the efficient heat removal from the typically 100\,$\mu$m-thick gain medium which is mounted on a water-cooled diamond substrate. These unique properties in combination with the reliability of industrial diode pumps makes Yb:YAG lasers potential drivers for high-energy, high-average power light transients. However, their narrowband emission cross-section \cite{Koerner2012}, in addition to the gain narrowing, limit their pulse duration to around 1\,ps at hundreds of mJ energy, and tens of ps at J energy \cite{nubbemeyer2017a,Baumgarten2016a}. Their long pulse duration makes the direct generation of multi-octave, CEP-stable pulses from Yb:YAG more challenging in comparison to Ti:Sa technology. 

In what follows we demonstrate an all-Yb:YAG multi-octave, frontend to pump and seed two OPCPA channels and their temporal compression to few-cycle pulses. We show that the intrinsic synchronization in this scheme reduces the temporal jitter only to long-term drift. The spatial and temporal stability of the synthesizer are confirmed by resolving the electric field of the free induction decay of water molecules with sub-cycle temporal resolution.

\section*{Results}

\subsection*{Mid-infrared and near-infrared OPCPAs}

A schematic of the frontend is depicted in Fig.\,1. A home-built, Kerr-lens mode-locked Yb:YAG thin-disk oscillator, delivering 2\,$\mu$J, 350\,fs pulses at 15\,MHz is used to seed a home-built, Yb:YAG thin-disk regenerative amplifier \cite{fattahi2016}. The amplifier operates in chirped-pulse amplification configuration. Therefore, the seed pulses are temporally stretched to 2\,ns in a pair of gold-gratings prior to amplification and compressed to their Fourier transform limit after the amplification. The laser delivers 20\,mJ, 1\,ps pulses at 5\,kHz repetition rates and serves as the only laser in the synthesizer.

1.8\,mJ of pump energy is used to generate a CEP-stable, multi-octave, broadband spectrum \cite{Fattahi2016c}. The resultant supercontinuum spans from 500\,nm to 2500\,nm (red curve in Fig.\,2-a) with 90\,mrad root mean squared CEP fluctuations over 100\,minutes measurement time. A broadband dielectric beam splitter \cite{amotchkina2016} is used to divide the seed spectrum into two portions: i) near-infrared (NIR) region from 700\,nm to 1400\,nm, and ii) mid-infrared (MIR) region with spectral coverage from 1600\,nm to 2500\,nm. The remaining energy of the Yb:YAG amplifier is frequency-doubled in a 0.5\,mm-thick Type I, barium borate (BBO) crystal at 80\,GW/cm$^2$ peak intensity. The second harmonic module delivers 9\,mJ pulses at 515\,nm. Thereafter, the second harmonic pulses are separated from the fundamental by a dielectric beam splitter.

A single-stage, non-collinear OPCPA containing a 4-mm-thick lithium triborate (LiB$_3$O$_5$, LBO) crystal is used to amplify the NIR region of the spectrum. 0.8\,mJ of the pump pulses at 515\,nm is focused to a 850\,$\mu$m beam diameter (full width at half maximum, FWHM), slightly larger than the 650\,$\mu$m (FWHM) seed beam. Pump and seed beams are crossed at an internal angle of 1.05$^\circ$ inside the LBO crystal with the phase-matching angle of $\phi = 15^o$ in Type I configuration. The amplified spectrum with 20\,$\mu$J of energy is shown in Fig.\,2-a (orange curve). Afterwards, the amplified beam is collimated and sent through a custom-designed chirped-mirror compressor. The compressor consists of 16 double-angle chirped mirrors with a group delay dispersion (GDD) of $-30\,fs^2$ per bounce, and a pair of thin fused-silica wedges at the Brewster angle for fine dispersion control. A second-harmonic frequency-resolved optical gating (SH-FROG) containing a 10-$\mu$m-thick BBO crystal is used to characterize the compressed pulses. Fig.\,2-b shows the measured and retrieved spectrogram, and the retrieved temporal intensity profile and phase of the compressed pulses to 6\,fs (FWHM).  

After separation by the dielectric beam splitter, the MIR seed pulses pass through an acousto-optic programmable dispersive filter (AOPDF) (Fastlite, Dazzler) and amplified in a single-stage degenerate OPCPA containing a 2\,mm-thick periodically-poled lithium niobate (PPLN) crystal with a poling period of 30.64\,$\mu$m. 180\,$\mu$J pulses at 1030\,nm with the peak intensity of 70\,GW/cm$^2$ are used to pump the MIR-OPCPA stage. 1$^\circ$ angle between the seed and pump beams is induced to facilitate the separation of the signal beam after amplification. Fig.\,2-a (brown curve) shows the spectrum of the 5\,$\mu$J amplified MIR pulses. The amplified pulses were compressed to 18\,fs by using a 2\,mm-thick silicon in combination with the introduced phase by Dazzler. Fig.\,2-c shows the measured and retrieved spectrograms, and the retrieved temporal intensity profile and phase of the MIR pulse. Theses measurements were done by a SH-FROG containing a 100-$\mu$m-thick BBO crystal.

\subsection*{Temporal and phase jitter}

Combining the MIR and NIR few-cycle pulses by using a broadband dielectric beam-combiner and their coherent superposition leads to the generation of light transients. In few-cycle coherent synthesis the relative temporal overlap between the few-cycle pulses results in a great variety of light transients \cite{Wendl2018}. Therefore, as with any kind of interferometer, relative phase and relative timing between the two arms of the synthesizer define the ultimate stability of the generated light transient \cite{Manzoni2015a, Rossi2017}. In our scheme, the CEP stability of the seed pulses results in the stability of the absolute phase of both OPCPA channel outputs. Moreover, the direct multi-octave seed generation from the pump laser assures the intrinsic temporal synchronization and restrict the temporal jitter only to long-term drift.
To avoid any temporal instability caused by fluctuations in the arrival of pump pulses in each OPCPA channel, the seed pulse duration was kept shorter than the pump pulse duration. This approach allows for decoupling the nonlinear amplification process from linear temporal jitter between the seed and pump pulses and restricts temporal jitter to long-term drifts caused by optical path of the synthesizer.

As the few-cycle pulses from both OPCPAs contain only two-three cycles of electric field, the temporal jitter between the two channels has to be lower than a fraction of their half-field cycle. To prove the intrinsic synchronization between the two channels over a time window of several picoseconds and restriction of the temporal jitter to long term drifts, we conducted an experiment with the new source.
 
Upon the interaction of broadband pulses with molecules, photons at the resonance frequencies are slowed down due to the increase in the refractive index and appear at the trailing edge of the excitation pulse. The delayed response, which is called free induction decay (FID) \cite{Brewer1972}, can be explained by Kramers-Kronig relation and lasts for several picoseconds in liquid phase. We aimed for the direct electric field detection of the MIR pulses and their FID after interaction with water molecules over several picoseconds by means of electro-optic sampling (EOS) \cite{Keiber2016}. The NIR pulses are used as probe pulses in EOS and are required to have a temporal jitter below the sub-cycle duration of the MIR pulses over the entire scanning range to allow for resolving the fast oscillating electric field of the MIR pulses and the FID. 

 \subsection*{Direct electric field sampling}
 
We chose to detect water's molecular response as water has a strong $\nu_2 + \nu_3$ combination band near 1930\,nm (5180\,cm$^{-1}$) \cite{Andrews2014}. The 18\,fs, MIR pulses were used to excite the combination band and were focused by a 4-inch focal length off-axis parabolic mirror into deionized water sandwiched between two 2.3\,mm-thick sapphire windows with 500\,$\mu$m spacing. Afterwards, the transmitted beam was collimated by another 4-inch focal length off-axis parabolic mirror and sent to the EOS setup for field detection (see Fig.\,3-a). 

The EOS setup is illustrated in Fig.\,3-b. The 6\,fs NIR pulses are collinearly combined with the transmitted MIR pulses from the sample in a broadband wire grid polarizer (Thorlabs, WP25L-UB) and focused into a 50\,$\mu$m-thick BBO crystal at the phase matching angle of $\theta=25^\circ$ for Type II sum frequency generation. The sum-frequency signal at 670\,nm has orthogonal polarization relative to the NIR probe pulses. To enhance the detection sensitivity, the spectral range between 650\,nm and 750\,nm of the transmitted beams after the BBO crystal is filtered out \cite{Porer2014}. The interference between the NIR probe pulses and the sum frequency signal is assured by a quarter wave plate. The resulted polarization rotation is read out by an ellipsometer consisting of a Wollaston prism and balanced photodiodes as a function of relative timing of the NIR and MIR pulses. The MIR pulses are modulated at 2.5\,kHz with a mechanical chopper, while the mechanical delay stage was tracked \cite{Schweinberger2019}.

Fig.\,3-c shows the measured electric field of the MIR pulses in the absence (dark blue curve) and presence (bright blue curve) of water. To probe the MIR pulses, the delay stage in the probe channel was scanned over 450\,$\mu$m with a speed of 1.7\,$\mu$m/s, corresponding to a single shot detection at each delay position. As can be seen, in the absence of water the amplitude of the excitation pulses goes to zero at temporal delays above 200\,fs. After the injection of water, the transmitted excitation pulse is temporally chirped and the molecular free induction decay is formed. Resolving the electric field of the MIR pulses with sub-cycle precision over 3\,ps time window and in the absence of active stabilization, demonstrates the spatial and temporal stability of the synthesizer. 

Moreover, here the molecular response can be detected free of background at temporal delays beyond 200\,fs, allowing for higher detection sensitivity. By employing broadband excitation pulses in combination with EOS the entire molecular vibrations in the fingerprint region can be simultaneously excited and detected, which holds promise for advancing femtosecond time-domain spectroscopy to direct measurement of the complex electric field and higher detection sensitivity. 

\section*{Discussion}

In summary, we demonstrated an all-ytterbium source for high-energy light transient generation based on OPCPA. The frontend covers a spectral range of 700\,nm to 2500\,nm and delivers intrinsically synchronized 6\,fs and 18\,fs pulses with the total energy of 25\,$\mu$J, at 5\,kHz repetition rates. The source is capable of delivering sub-cycle light transients with 3\,fs temporal duration. The stable CEP was ensured by generating passively CEP-stable multi-octave seed pulses directly from the Yb:YAG amplifier. The temporal jitter between the NIR and MIR pulses in this scheme is only dominated by long-term drift. 

As a first application, we demonstrated the direct measurement of the transmitted electric field of the MIR pulses after interaction with water with sub-cycle resolution. The linear absorption of the MIR pulses by water molecules forms a fee induction decay at the trailing edge of the MIR pulses that lasts for several ps. Few-cycle, NIR pulses were used to probe the oscillating MIR electric field over a 3\,ps time window, which proves the relative spatial and temporal stability of the MIR and NIR channels. Moreover, this is the first measurement of the complex electric field of the FID of overtone and combination bands to the best of our knowledge. The demonstrated control and measurement of the electric field of light at 0.5\,PHz holds promise to open up new opportunities for precise observation and control of molecular vibrations over the entire molecular finger print region down to a few femtosecond time scales. 

Multi-mJ, light transients hold promise for advancing attosecond pulse generation to higher power level and higher photon energies. Our simulations suggest that by using the available pump energy in additional amplifier stages the energy of the two-octave spectrum can be scaled to 3.8\,mJ (Fig.\,4). Higher energy and average power can be achieved by employing a 5\,kW, 200\,mJ Yb:YAG amplifier \cite{nubbemeyer2017a} as the driving laser.

\section*{Materials and Methods}
The presented three-dimensional simulations in Fig.\,4 were performed by SISYFOS \cite{Arisholm1997}. In this code, optical parametric frequency conversions in nonlinear media are simulated by utilizing a Fourier‐space method. SISYFOS solves the coupled differential equations for the slowly varying amplitudes. This model can take into account most of the relevant physical effects in the frequency conversion such as propagation effects, second and third order nonlinearity, thermal effects, two-photon absorption, and non-collinear interactions. In this numerical simulation, we neglected the thermal effects, third-order nonlinear processes, and two photon absorption. The Sellmeier coefficients are obtained from \cite{Kato1994a,Zelmon1997}. 
For simulating the NIR channel, the amplified spectrum in the first stage (orange curve in Fig.\,2-a) was further amplified into two additional amplification stages. The remaining energy (8.3\,mJ) from the second harmonic module at 515\,nm was used to pump the second stage and recycled at the third amplification stage. The pump’s beam size was adjusted to keep its peak intensity at 100 GW/cm$^2$. The parametric amplification was formed in an LBO crystal with a phase-matching angle of 15.5$^o$, an internal non-collinear angle of 1.05$^o$, and $d_{eff} = 0.82 pm/V$ resulting in the amplified pulse energy of 1.7\,mJ. The detailed parameters of the NIR chain are summarized in Table \ref{tab:1}.

For simulating the amplified spectrum of the MIR channel (brown curve in Fig.\,2-a), 8.92\,mJ of pump energy at 1030\,nm was used to pump two OPCPA stages, while the pump energy was recycled at the last stage. The pump intensity in all stages is set to 70 GW/cm$^2$. The signal energy was boosted to 2.1\,mJ in the MIR chain by using a type‐I, lithium niobate (LiNbO$_3$) crystal with a phase‐matching angle $\theta= 42.9^o$ and assumed d$_{eff}$ value of 3.96\,pm/V. Table \ref{tab:2} summarizes the simulation parameters of the MIR channel.

\begin{table}[h]
    \centering
    \caption{Simulation parameters of the NIR channel. L$_c$: crystal thickness, E$_p$: input pump energy, E$_s$: amplified signal energy, $\omega_{p}$: pump beam radius (FWHM), $\omega_{s}$: signal beam radius (FWHM).}
    \begin{tabular}{|c |c |c |c |c |c |c|}
    \hline
	stage & L$_c$ & E$_p$ & E$_s$ & $\omega_{p}$ & $\omega_{s}$ & efficiency\\
          & (mm) & (mJ) & (mJ) & (mm) & (mm) & (\%)\\
    \hline
	$2^{nd}$  & 2.5 & 8.3 & 1.3 & 1.46 & 1.16 & 15.1\\
    \hline
	$3^{rd}$  & 1.0 & 6.0 & 1.8 & 1.28 & 1.02 & 9.8\\
    \hline

    \end{tabular}
    \label{tab:1}
\end{table}
\begin{table}[h]
    \centering
    \caption{Simulation parameters of the MIR channel. L$_c$: crystal thickness, E$_p$: input pump energy, E$_s$: amplified signal energy, $\omega_{p}$: pump beam radius (FWHM), $\omega_{s}$: signal beam radius (FWHM).}
    \begin{tabular}{|c |c |c |c |c |c |c|}
    \hline
	stage & L$_c$ & E$_p$ & E$_s$ & $\omega_{p}$ & $\omega_{s}$ & efficiency\\
          & (mm) & (mJ) & (mJ) & (mm) & (mm) & (\%)\\
    \hline
	$2^{nd}$  & 2.0 & 8.9 & 1.3 & 1.63 & 1.50 & 14.9\\
    \hline
	$3^{rd}$  & 1.0 & 6.2 & 2.2 & 1.42 & 1.34 & 13.9\\
    \hline
    \end{tabular}
    \label{tab:2}
\end{table}



\bibliography{library}

\bibliographystyle{Science}

\section*{Acknowledgments}
H.F. acknowledges the MINERVA Scholarship of Max Planck Society. A.A. acknowledges scholarship by King Saud University, Deanship of Scientific Research, College of Science Research Center. All data needed to evaluate the conclusions in the paper are present in the paper and/or the Supplementary Materials. Additional data related to this paper may be requested from the authors. H.F., and F.K. designed and initiated the study; Experiments were performed by A.A., H.W., and H.F.; G.B., N.A., S.A.H., and W.S have contributed in the development of the setup; Design and manufacturing of the optics were carried out by V.P.; The manuscript was written by H.F., A.A. and H.W.; all authors discussed the results and contributed to the manuscript.
\section*{Competing Interests} 
The authors declare that they have no competing interests.
\clearpage
\begin{figure}[t]
\centerline{\includegraphics[width=0.9\columnwidth]{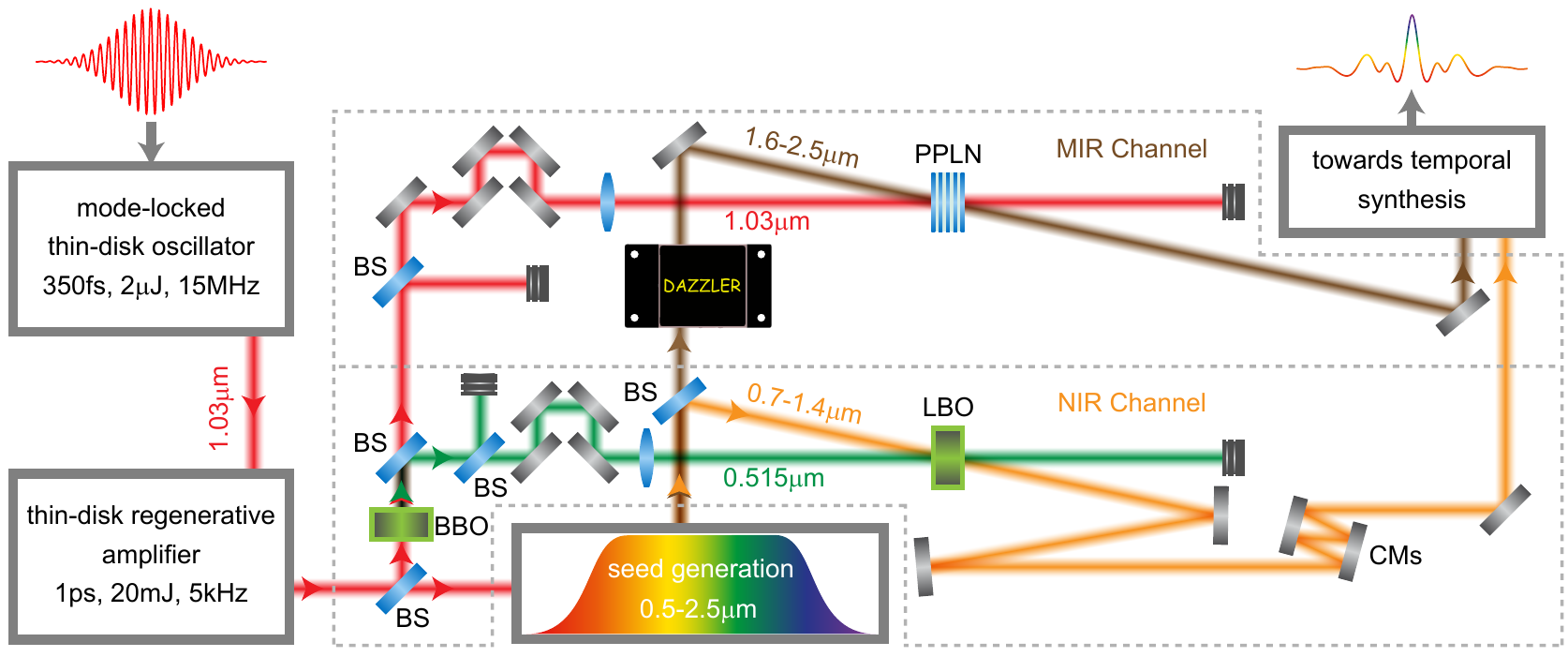}}
\caption{All-ytterbium frontend for high average-, and peak-power light transient generation. The setup is based on a Yb:YAG thin-disk amplifier, seeded by a Kerr-lens mode-locked thin-disk oscillator. 1.8\,mJ of the laser energy is used to directly generate mulit-octave, CEP-stable seed pulses. Thereafter, the broadband spectrum is divided into two channels, near-infrared channel (NIR) spanning from 700\,nm to 1400\,nm, and mid-infrared channel (MIR) covering 1600\,nm-2500\,nm. Each channel is amplified in a single stage OPCPA pumped by fundamental or second harmonic of the laser and compressed to its Fourier limit. The  energy of few-cycle pulses can be increased by adding consecutive OPCPA stages in each channel. Coherent synthesis of the two few-cycle pulses results in generation of light transients. BS: beam splitter, CMs: chirped mirrors.}
\label{fig:1}
\end{figure}
\begin{figure}[t]
\centerline{\includegraphics[width=0.8\columnwidth]{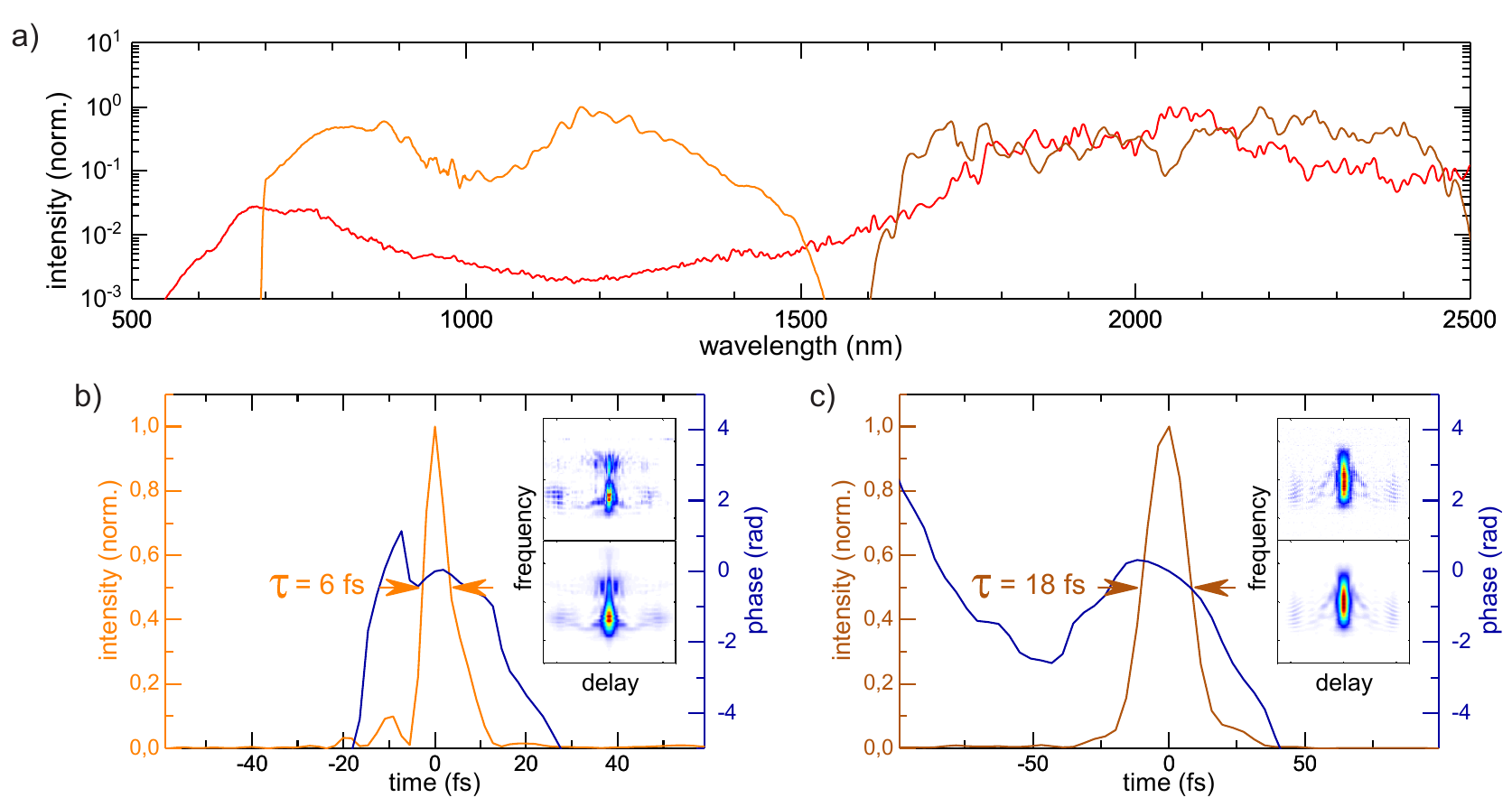}}
\caption{The laser specifications. a) CEP-stable, multi-octave spectrum of seed pulses (red). Amplified spectra of NIR (orange) and MIR (brown) OPCPAs. Retrieved temporal intensity profile and phase of the compressed pulses in b) NIR and c) MIR channels. Inset: measured (top) and retrieved (bottom) FROG spectrograms.}
\label{fig:2}
\end{figure}
\begin{figure}[t]
\centerline{\includegraphics[width=0.9\columnwidth]{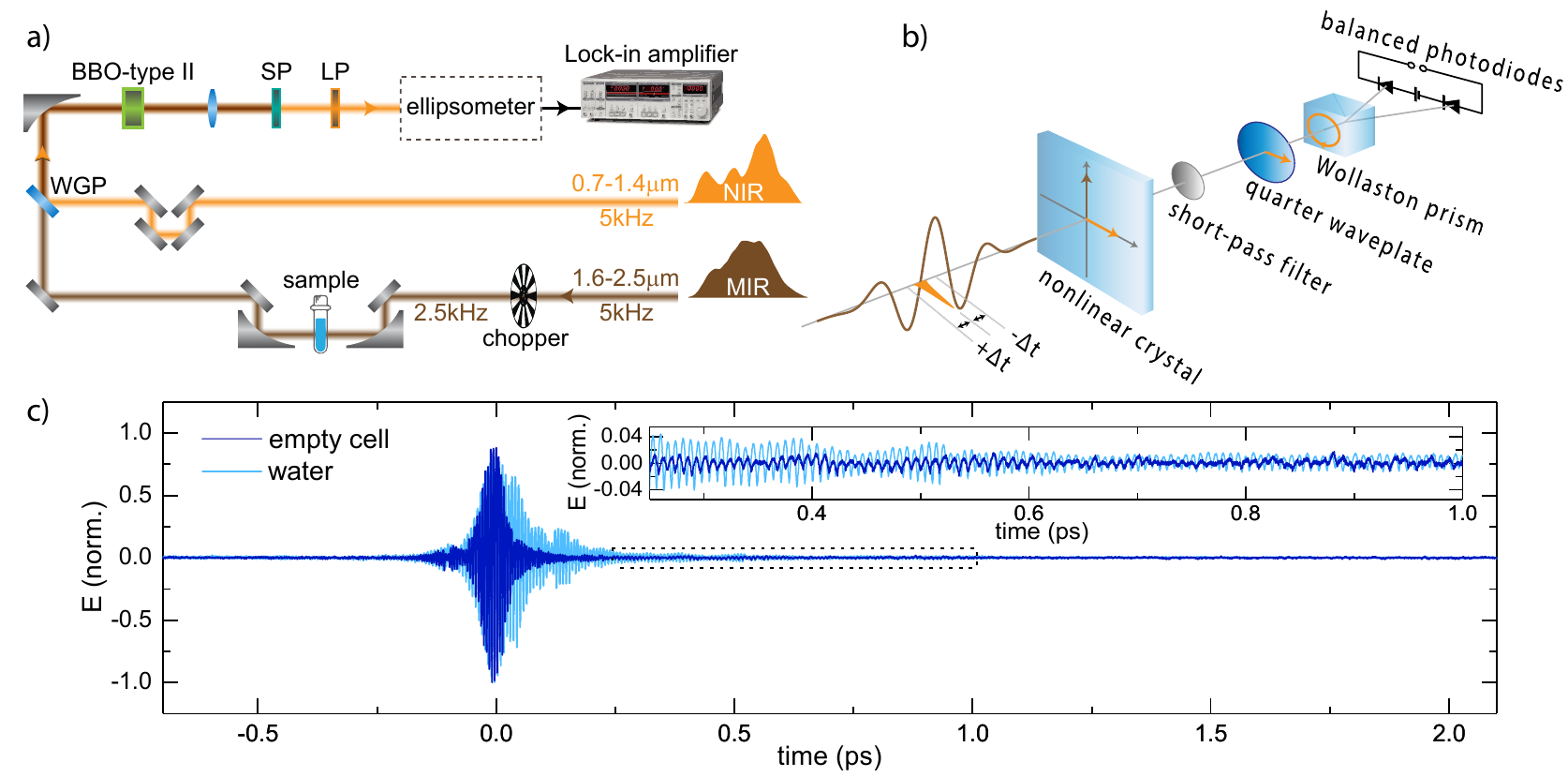}}
\caption{Direct electric field detection. a) Schematic of the linear absorption setup for generating and sampling the FID of water molecules. b) EOS setup containing a 50\,$\mu$m-thick BBO (Type II) crystal and an ellipsometer. The MIR pulses are chopped at 2.5\,kHz. The NIR beam after the delay line is collinearly combined with the MIR beam and both are focused in the EOS crystal for sum-frequency generation. The generated sum-frequency signal spectrally overlaps and temporally interferes with the high-frequency components of the probe beam. Appropriate spectral filtering is used to enhance the EOS signal. The polarization rotation as a function of time delay is detected by an ellipsometer and read out by a lock-in amplifier. c) The measured electric field of the MIR pulses in the absence (dark blue curve) and the presence of water (light blue curve). Inset: zoomed plot in a shorter temporal window. WGP: wire grid polarizer, SP: short-pass filter, LP: long-pass filter.}
\label{fig:3}
\end{figure}
\begin{figure}[t]
\centerline{\includegraphics[width=0.9\columnwidth]{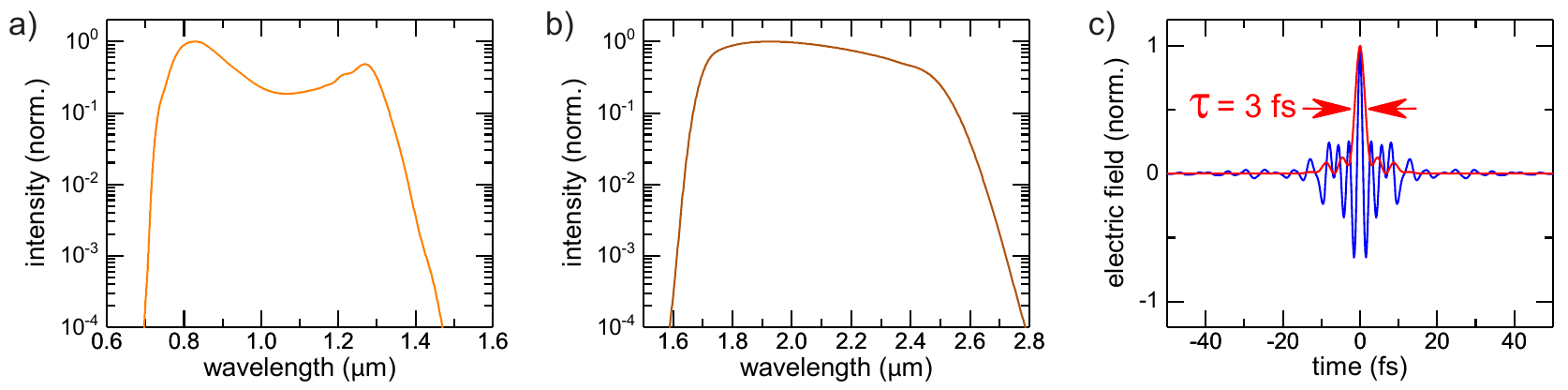}}
\caption{Simulated amplified spectra in a) NIR channel with 2\,mJ of energy and b) MIR with 2\,mJ of energy and c) their calculated synthesized waveform.}
\label{fig:4}
\end{figure}
\end{document}